\documentclass{PoS}

\usepackage{amsmath}
\usepackage{bm}
\usepackage{psfrag}

\newcommand*{\bea}{\begin{eqnarray}}
\newcommand*{\eea}{\end{eqnarray}}

\newcommand*{\chpt}{$\chi$PT}

\def\spose#1{\hbox to 0pt{#1\hss}}
\def\ltapprox{\mathrel{\spose{\lower 3pt\hbox{$\mathchar"218$}}
 \raise 2.0pt\hbox{$\mathchar"13C$}}}
\def\gtapprox{\mathrel{\spose{\lower 3pt\hbox{$\mathchar"218$}}
 \raise 2.0pt\hbox{$\mathchar"13E$}}}
\def\inapprox{\mathrel{\spose{\lower 3pt\hbox{$\mathchar"218$}}
 \raise 2.0pt\hbox{$\mathchar"232$}}}

\title{Light pseudoscalar meson masses and decay constants from mixed action lattice QCD}

\ShortTitle{Mixed action decay constants}

\author{Christopher Aubin \\ 
	Department of Physics, College of William and Mary, Williamsburg, VA, USA\\
		E-mail: \email{caaubin@wm.edu}}

\author{\speaker{Jack Laiho}\\
        Department of Physics, Washington University, St. Louis, MO, USA\\
        E-mail: \email{jlaiho@fnal.gov}
       }
       
\author{Ruth S. Van de Water\\
Theoretical Physics Department, Fermi National Accelerator Laboratory, Batavia, IL, USA\\
E-mail: \email{ruthv@bnl.gov}}

\abstract{We calculate the light pseudoscalar decay constants, $f_\pi$ and $f_K$, and their ratio using domain wall valence quarks and 2+1 flavors of dynamical staggered quarks.  Use of the MILC gauge configurations allows us to simulate at several sea quark masses and spatial volumes, and with two lattice spacings.  We study how well our numerical lattice data for light decay constants and meson masses is described by next-to-leading order $SU(3)$ mixed action chiral perturbation theory and explain our strategy for the chiral and continuum extrapolation. Combining our result for $f_K/f_\pi$ with experimental measurements of pion and kaon leptonic decays allows a model-independent determination of $|V_{us}|/|V_{ud}|$;  we find a preliminary value of $|V_{us}|/|V_{ud}| = 0.2315(45)(7)$.}

\FullConference{The XXVI International Symposium on Lattice Field Theory\\
		 July 14-19 2008\\
		 Williamsburg, Virginia, USA}

\begin{document}

\section{Introduction}

We calculate the masses and decay constants of light pseudoscalar mesons in unquenched lattice QCD.  These quantities serve as benchmarks for lattice calculations because they are relatively simple to compute with all sources of systematic errors under control.  They also provide a test of chiral perturbation theory ($\chi$PT), including the more sophisticated versions that include lattice discretization effects.  Successful calculation of these simple quantities lends confidence to calculations of more complicated quantities such as the kaon bag parameter ($B_K$) and $K\to\pi\pi$ matrix elements, which also rely on $\chi$PT-guided extrapolations.  Furthermore, the ratio of decay constants $f_K/f_\pi$ is interesting in its own right because it allows for a model-independent determination of the ratio of CKM matrix elements $|V_{us}|/|V_{ud}|$~\cite{Marciano:2004uf, Aubin:2004fs}.  


Following the approach of the LHP Collaboration, we use HYP-smeared domain-wall valence quarks and staggered sea quarks \cite{Renner:2004ck}.  We use gauge configurations generated by the MILC Collaboration with a 2+1 flavor improved staggered action because they are publicly available and span a wide range of quark masses, lattice spacings, and volumes~\cite{Aubin:2004fs}.  These lattices allow for good control over chiral and continuum extrapolations for light pseudoscalar quantities using staggered valence quarks.  We show in this work that we have similar control over these extrapolations using domain-wall valence quarks.  We account for the effects of staggered sea quarks by using the appropriate mixed action $\chi$PT (MA$\chi$PT) expressions~\cite{Bar:2005tu}.  The $\chi$PT formulas for light pseudoscalar quantities in the mixed-action theory are more continuum-like than in the purely staggered case, with fewer new parameters.  The mixed-action approach is especially powerful when considering more complicated quantities that are not protected from operator mixing under renormalization, such as $B_K$.  In the staggered case, mixing with taste-breaking operators under renormalization creates significant complications~\cite{VandeWater:2005uq}.  In the mixed-action case, the chiral symmetry of the valence sector makes nonperturbative renormalization as straightforward as in dynamical domain-wall simulations~\cite{Aubin:2007pt}.   The results presented in this work bolster confidence in our future mixed-action calculation of $B_K$.

In Ref.~\cite{Aubin:2008wk} we have performed a strong check of the ability of MA\chpt\ to accurately describe  discretization effects by investigating the isovector scalar correlator.  We find that the MA\chpt\ prediction for the two-particle intermediate state (bubble) contribution to the scalar correlator is in good quantitative agreement with the numerical lattice data, even though there are large discretization effects due to staggered sea quarks.  Thus we conclude that MA\chpt\ correctly describes the dominant unitarity-violating contributions to mixed-action lattice simulations.  Fortunately, in the case of most weak-matrix elements, MA\chpt\ predicts that non-analytic unitarity-violating errors should contribute at the sub-percent level on the MILC lattices that we are using.  This fact, in conjunction with our successful analysis of the scalar correlator, substantiates the claim that unitarity-violating effects in mixed-action lattice simulations can be accounted for and removed to recover precise continuum values for weak matrix elements.

\section{Lattice calculation and chiral-continuum extrapolation}

\begin{table}
\begin{center}
\caption{Simulation parameters of MILC staggered gauge configurations used in this work.  ``Nominal" quark masses are shown.}
 \label{tab:params}
\begin{tabular}{ccccccc}
  \hline\hline
  $a$(fm) & $a\hat{m}' / am'_s$ & $L$(fm) & $m_\pi L$ & $10/g^2$ & Lat. Dim. & $\#$ Confs.  \\
  \hline
  $\approx 0.12$ & $0.02/0.05$ & 2.4 & 6.2 & 6.79 & $20^3\times 64$ & 117  \\
  $\approx 0.12$ & $0.01/0.05$ & 2.4 & 4.5 & 6.76 & $20^3\times 64$ & 220  \\
  $\approx 0.12$ & $0.007/0.05$ & 2.4 & 3.8 & 6.76 & $20^3\times 64$ & 268  \\
  $\approx 0.12$ & $0.005/0.05$ & 2.9 & 3.8 & 6.76 & $24^3\times 64$ & 216  \\
  $\approx 0.12$ & $0.01/0.03$ & 2.4 & 4.5 & 6.76 & $20^3\times 64$ & 160  \\
  \hline
  $\approx 0.09$ & $0.0124/0.031$ & 2.4 & 5.8 & 7.11 & $28^3\times 96$ & 198  \\
  $\approx 0.09$ & $0.0062/0.031$ & 2.4 & 4.1 & 7.09 & $28^3\times 96$ & 210  \\
  $\approx 0.09$ & $0.0031/0.031$ & 3.4 & 4.2 & 7.08 & $40^3\times 96$ & 38  \\
  $\approx 0.09$ & $0.0062/0.0186$ & 2.4 & 4.1 & 7.10 & $28^3\times 96$ & 160  \\
  \hline\hline
\end{tabular}
\end{center}
\vspace{-7mm}
\end{table}

We generate data on the MILC ensembles given in Table \ref{tab:params}.  The quantities $m'_s$ and $\hat{m}'$ denote the values of the simulated staggered sea quark masses, while the unprimed quantities denote the physical masses $m_s$ and $\hat{m}=(m_u+m_d)/2$.  We compute domain-wall valence quark propagators with masses from $m_s/10 - m_s$; our lightest pion is $\approx 240$ MeV.  In order to minimize finite-volume effects, we restrict the combination $m_\pi L \gtapprox 4$.  We do not tune the masses of our valence-valence pions to any particular values because one cannot recover unitarity to obtain full QCD at nonzero lattice spacing in the mixed-action theory.  We instead generate many partially quenched data points at two lattice spacings, and use MA$\chi$PT to extrapolate to the physical quark masses and continuum.  The one-loop MA$\chi$PT expressions for most quantities of interest are the same as for chiral fermions except for the appearance of additive shifts to the sea-sea and valence-sea squared meson masses.  Because the two new splittings are easy to obtain from spectrum calculations~\cite{Aubin:2004fs,Aubin:2008wk}, they do not need to be included as free parameters in chiral fits.


In order to suppress contamination from pions circling the lattice in the time direction, we use symmetric and antisymmetric linear combinations of quark propagators with periodic and antiperiodic boundary conditions in our 2-point correlation functions.  The approximate chiral symmetry of domain-wall quarks allows us to use the pseudoscalar current to obtain $f_\pi$, and we extract the decay constant using the axial Ward identity: 
	\bea\label{eq:fpi} f_P = \frac{A_{WP}}{\sqrt{A_{WW}}}\frac{\sqrt{2}(m_x+m_y+2 m_{res})}{m_\pi^{3/2}}, 
	\eea
\noindent where $A_{WW}$, $A_{WP}$, and $m_\pi$ come from a simultaneous fit of Coulomb gauge-fixed wall-wall and wall-point correlators.    This channel is statistically cleaner than the axial current, and is protected by a non-renormalization theorem.  Because chiral symmetry is not exact in our simulations at finite $L_S=16$, there are small corrections to the axial Ward identity.   We account for the dominant corrections due to residual chiral symmetry breaking by including $m_{res}$ in Eq.~(\ref{eq:fpi}).  We estimate the subleading ${\cal O}(m_{res}m_q a^2)$ corrections to the axial Ward identity using our results at two lattice spacings and largely remove them in the continuum extrapolation.

%
%

\begin{figure}
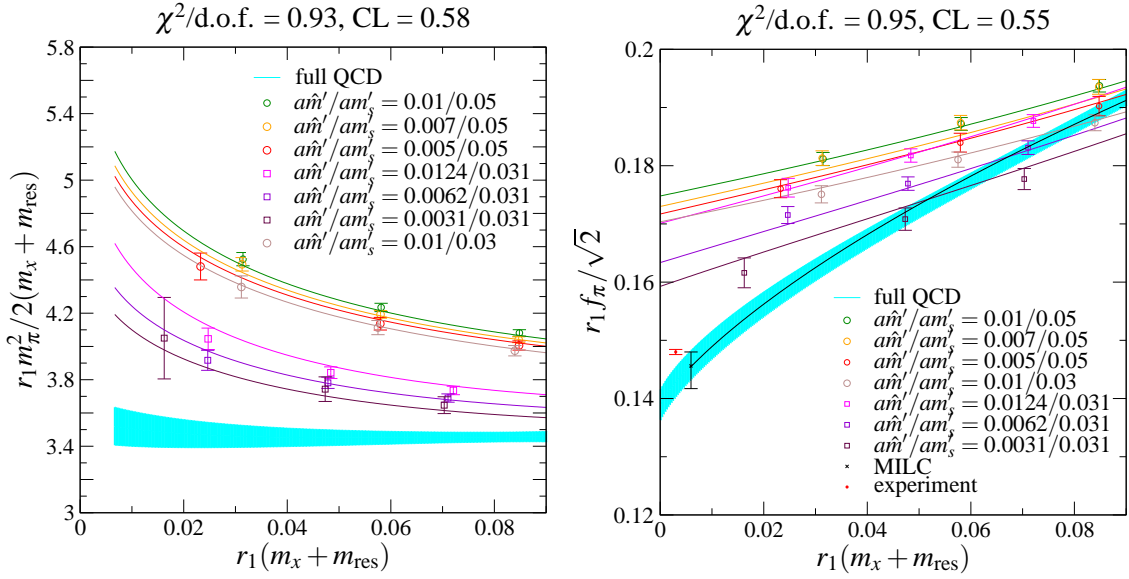

\bigskip
\begin{center}
\vspace{-4mm}
\begin{tabular}{c c}
	\psfrag{title}[b][b][1.0]{$\chi^2$/d.o.f. = 0.93, CL = 0.58}
	\psfrag{x-axis}[t][t][1.0]{$r_1 (m_x + m_\text{res}) $}
	\psfrag{y-axis}[b][b][1.0]{$r_1 m_\pi^2 / 2 (m_x + m_\text{res})$}
	\psfrag{full QCD}[b][b][0.8]{full QCD}
	\psfrag{01/05}[l][l][0.8]{$a\hat{m}' / am_s' = 0.01/0.05$}
	\psfrag{007/05}[l][l][0.8]{$a\hat{m}' / am_s' = 0.007/0.05$}
	\psfrag{005/05}[l][l][0.8]{$a\hat{m}' / am_s' = 0.005/0.05$}
	\psfrag{0124/031}[l][l][0.8]{$a\hat{m}' / am_s' = 0.0124/0.031$}
	\psfrag{0062/031}[l][l][0.8]{$a\hat{m}' / am_s' = 0.0062/0.031$}
	\psfrag{0031/031}[l][l][0.8]{$a\hat{m}' / am_s' = 0.0031/0.031$}
	\psfrag{01/03}[l][l][0.8]{$a\hat{m}' / am_s' = 0.01/0.03$}
	\includegraphics[scale=.41]{mpi_05_2008_psfrag.eps} &
	\psfrag{title}[b][b][1.0]{$\chi^2$/d.o.f. = 0.95, CL = 0.55}
	\psfrag{x-axis}[t][t][1.0]{$r_1 (m_x + m_\text{res}) $}
	\psfrag{y-axis}[b][b][1.0]{$r_1 f_\pi / \sqrt{2}$}
	\psfrag{MILC}[l][l][0.8]{MILC}
	\psfrag{exp}[l][l][0.8]{experiment}
	\psfrag{full QCD}[l][l][0.8]{full QCD}
	\psfrag{01/05}[l][l][0.8]{$a\hat{m}' / am_s' = 0.01/0.05$}
	\psfrag{007/05}[l][l][0.8]{$a\hat{m}' / am_s' = 0.007/0.05$}
	\psfrag{005/05}[l][l][0.8]{$a\hat{m}' / am_s' = 0.005/0.05$}
	\psfrag{0124/031}[l][l][0.8]{$a\hat{m}' / am_s' = 0.0124/0.031$}
	\psfrag{0062/031}[l][l][0.8]{$a\hat{m}' / am_s' = 0.0062/0.031$}
	\psfrag{0031/031}[l][l][0.8]{$a\hat{m}' / am_s' = 0.0031/0.031$}
	\psfrag{01/03}[l][l][0.8]{$a\hat{m}' / am_s' = 0.01/0.03$}
	\includegraphics[scale=.41]{fpi_07_wMILC_psfrag.eps}
\end{tabular}
\caption{Light pseudoscalar meson mass-squared (left plot) and decay constant (right plot) versus valence quark mass.  Full QCD curves are obtained using the ratio of the quark normalization factors from tree-level $\chi$PT to convert staggered quark masses to domain wall masses in the MA$\chi$PT formulas.  Only degenerate ($m_x=m_y$) points are shown. \label{fig:light}}
\end{center}
\vspace{-7mm}
\end{figure}

In order to convert lattice quantities into physical units we use the MILC Collaboration's determination of the scale, $r_1$, where $r_1$ is related to the force between static quarks, $r_1^2F(r_1)=1.0$~\cite{Sommer:1993ce,Bernard:2000gd}.  The ratio $r_1/a$ can be calculated precisely on each ensemble from the static quark potential.  We use the mass-independent prescription for $r_1$ described in Ref.~\cite{Bernard:2007ps}.  In order to fix the absolute lattice scale, one must compute a physical quantity that can be compared directly to experiment; we use the $\Upsilon$ 2S--1S splitting \cite{Gray:2005ad} and the most recent MILC determination of $f_\pi$ \cite{Bernard:2007ps}.  The combination of the $\Upsilon$ mass-splitting and the continuum-extrapolated $r_1$ value at physical quark masses leads to the determination $r_1^{\rm phys}=0.318(7)$ fm~\cite{Bernard:2005ei}.  The use of $f_\pi$  to set the scale yields $r_1^{\rm phys}=0.3108(15)(^{+26}_{-79})$ fm \cite{Bernard:2007ps}.  This difference between the two scale determinations leads to a systematic error in our decay constants labeled ``input $r_1$'' in Table~\ref{tbl:error}.


We use the $SU(3)$ MA$\chi$PT formulas derived in Ref.~\cite{Bar:2005tu} to extrapolate our numerical lattice data to the continuum and physical quark masses.  The choice of $SU(3)$ $\chi$PT is appropriate given the parameters of our numerical simulations because our light pion masses range from 240-500 MeV and are not much lighter than the physical kaon, which is integrated out in $SU(2)$ $\chi$PT. Furthermore, the largest of the taste-splittings on the coarse lattices is not much smaller than the kaon mass [$a^2\Delta_I\approx (460)^2 \textrm{MeV}^2$], though on the fine lattices it is about a factor of 3 times smaller [$a^2\Delta_I \approx (280)^2 \textrm{MeV}^2$].  The statistical errors on $m_P$ and $f_P$ are $\sim 0.5\%-2\%$ for most of our data points.  It is now well-established that NLO $\chi$PT does not describe pseudoscalar masses and decay constants to percent-level accuracy at the physical kaon mass, nor is it expected to based on power counting.  Our data set confirms this picture.  In order to get good fits (as measured by the correlated $\chi^2/$d.o.f.) to even our low-mass data we must include NNLO analytic terms.  The two-loop NNLO logarithmic corrections to the partially-quenched $\chi$PT have been computed in Ref.~\cite{Bijnens:2006jv}, but we have not yet implemented these terms in our fits because of their extremely complicated form.  These expressions would also have to be modified to account for the staggered sea sector, though, given our experience with the one-loop modifications to the mixed action, this is likely a small effect.  In the region where the NNLO analytic terms that we have added are important, we expect the NNLO logarithms to vary slowly enough that their effect is well approximated by the analytic terms.  Even so, this somewhat ad hoc treatment gives rise to our largest systematic error in the determination of decay constants.

We perform chiral fits using two different mass ranges: one with all pions less than 500 MeV in order to study the chiral expansion, and another including masses up to the strange quark, which is used to determine our central values for masses and decay constants.  We correct all data points for finite-volume effects using one-loop MA$\chi$PT, and take as a conservative estimate for the residual finite-volume errors the entire one-loop correction at the data point where that correction is largest.  The light-mass fits to $m_P^2/2(m_x+m_{\rm res})$ and to $f_P$ are shown in Figure~\ref{fig:light}.  For the leading-order low-energy constants (LEC's) appearing in the chiral Lagrangian we use the physical values ($f_\pi$, $\mu$), rather than the values obtained in the $SU(3)$ chiral limit.  To one-loop order this is consistent, and studies by both the MILC and JLQCD Collaborations suggest that use of a physical parameter for the chiral coupling ($f_\pi$ rather than $f_0$) is likely to result in a more convergent chiral expansion~\cite{Aubin:2004fs,Noaki:2008iy}.  We fit the pseudoscalar masses and decay constants separately, since they have no common parameters through one-loop once the leading order LEC's are set to the physical values.  We vary the parameter $f$ appearing in the one-loop expression between the values $f_0$ and $f_K$ as part of our estimate of the chiral fit systematic error.  Each of our low-mass fits has 42 data points and 9-10 fit parameters; our high-mass fits have 69 data points and 15-16 fit parameters.  We do not constrain the size of the LEC's with priors.  Figure~\ref{fig:light} shows the continuum full QCD curve for $f_\pi$ with statistical errors as a cyan band.  The black curve on the decay constant plot is the continuum full QCD result determined by MILC using staggered fermions, after rescaling the bare quark mass in the $x$-axis by the appropriate factor.  Despite the very different shapes of the partially quenched mixed-action data and staggered data (not shown; see Ref.~\cite{Aubin:2004fs}) at finite lattice spacing, the continuum curves are in good agreement.  We also show the experimentally-measured value of $f_\pi$ (which is offset  for clarity) using the $\Upsilon$ spectrum to set the scale for comparison.

\begin{figure}
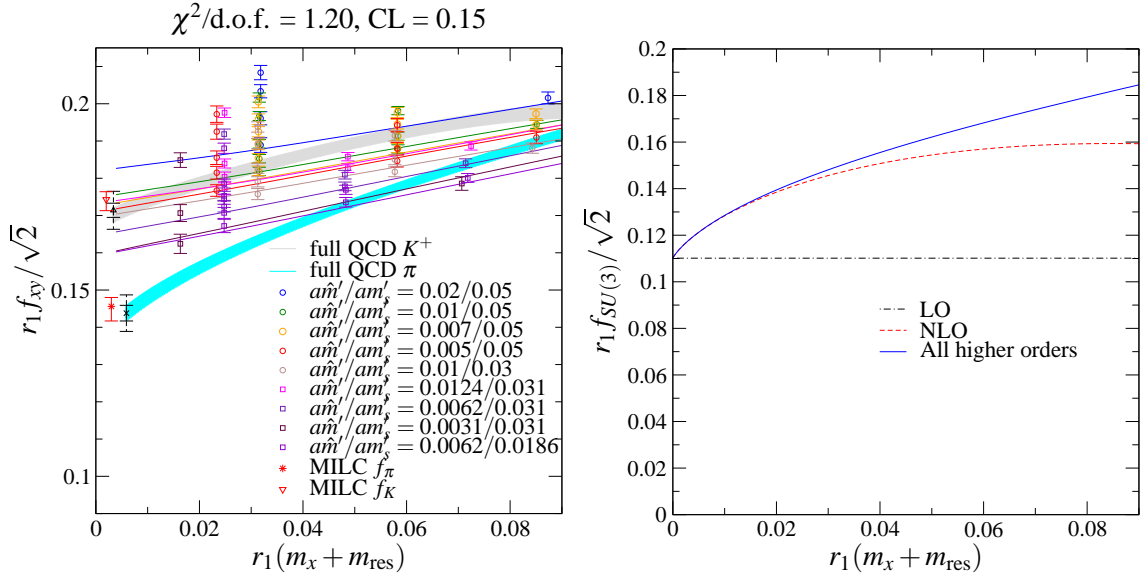

\bigskip
\begin{center}
\vspace{-4mm}
\begin{tabular}{c c} 
	\psfrag{title}[b][b][1.0]{$\chi^2$/d.o.f. = 1.20, CL = 0.15}
	\psfrag{x-axis}[t][t][1.0]{$r_1 (m_x + m_\text{res})$}
	\psfrag{y-axis}[b][b][1.0]{$r_1 f_{xy} / \sqrt{2}$}
	\psfrag{MILC fpi}[l][l][0.8]{MILC $f_\pi$}
	\psfrag{MILC fK}[l][l][0.8]{MILC $f_K$}
	\psfrag{full QCD K}[l][l][0.8]{full QCD $K^+$}
	\psfrag{full QCD pi}[l][l][0.8]{full QCD $\pi$}
	\psfrag{02/05}[l][l][0.8]{$a\hat{m}' / am_s' = 0.02/0.05$}
	\psfrag{01/05}[l][l][0.8]{$a\hat{m}' / am_s' = 0.01/0.05$}
	\psfrag{007/05}[l][l][0.8]{$a\hat{m}' / am_s' = 0.007/0.05$}
	\psfrag{005/05}[l][l][0.8]{$a\hat{m}' / am_s' = 0.005/0.05$}
	\psfrag{0124/031}[l][l][0.8]{$a\hat{m}' / am_s' = 0.0124/0.031$}
	\psfrag{0062/031}[l][l][0.8]{$a\hat{m}' / am_s' = 0.0062/0.031$}
	\psfrag{0031/031}[l][l][0.8]{$a\hat{m}' / am_s' = 0.0031/0.031$}
	\psfrag{01/03}[l][l][0.8]{$a\hat{m}' / am_s' = 0.01/0.03$}
	\psfrag{0062/0186}[l][l][0.8]{$a\hat{m}' / am_s' = 0.0062/0.0186$}
	\includegraphics[scale=.41]{fk_all_10_2008_psfrag.eps} &
	\psfrag{x-axis}[t][t][1.0]{$r_1 (m_x + m_\text{res}) $}
	\psfrag{y-axis}[b][b][1.0]{$r_1 f_{SU(3)} / \sqrt{2}$}
	\psfrag{LO}[l][l][0.8]{LO}
	\psfrag{NLO}[l][l][0.8]{NLO}
	\psfrag{All orders}[l][l][0.8]{All higher orders}
	\includegraphics[scale=.41]{fpi_NLO_psfrag.eps}
\end{tabular}
\caption{High-mass fit to $f_P$ data with chiral continuum extrapolation curves for $f_K$ and $f_\pi$ on left.  For nondegenerate mass points, the lighter valence quark mass is shown along the $x$-axis.  Again, full QCD curves are obtained using the ratio of the quark normalization factors from tree-level $\chi$PT to convert staggered quark masses to domain wall masses in the MA$\chi$PT formulas.  Study of convergence of $\chi$PT for decay constants on right. \label{fig:fk}}
\end{center}
\vspace{-7mm}
\end{figure}

We present the result of the high-mass fit used to obtain $f_\pi$ and $f_K$ in Figure~\ref{fig:fk}.  The cyan and grey bands show the continuum full QCD extrapolations with statistical errors for degenerate ($f_\pi$) and nondegenerate ($f_K$) decay constants, respectively.  For the nondegenerate curve, the strange mass is fixed at its physical value.  We determine the physical bare valence $u$-, $d$-, and $s$-quark masses from our fits to $m_P^2$; the physical bare sea quark masses were determined by MILC in a similar manner.  In order to obtain the charged pion and kaon decay constants, we extrapolate the light valence-quark mass to $\hat{m}$ for $f_\pi$ but to $m_u$ for $f_K$.  The light sea-quark masses are set to $\hat{m}$ for both decay constants, and this leads to a slight vertical offset in the final value for $f_K$ as compared to the grey full QCD band in Figure~\ref{fig:fk}.  In order to achieve good fits including nondegenerate pseudoscalar masses up to $\approx 600$ MeV, we must include higher-order polynomial terms.  Figure~\ref{fig:fk} shows all of the data points used in the high-mass $f_P$ fit.  In addition to the NNLO analytic terms, we include terms proportional to $(m_x+m_y)^n$. We find that we cannot get good correlated fits unless $n$ is as high as six.  Figure~\ref{fig:fk} compares our results for decay constants with those obtained by the MILC Collaboration.  This comparison is more meaningful than a comparison with experiment because we are using the MILC $r_1$ (obtained from $f_\pi$) to set the scale, and one needs $|V_{us}|$ in order to extract $f_K$ from experiment.  Figure~\ref{fig:fk} also shows the breakdown of our high-mass fit including terms through LO, NLO, and ``all higher orders'' to the degenerate SU(3) curve in the continuum limit.  The right-most part of this plot corresponds to $500$ MeV, where we do not expect $\chi$PT to be especially convergent.  Because we are interpolating in the quark mass in this region, we expect the high-order polynomials to approximate higher-order nonanalytic effects, whereas closer to the physical pion mass, the NLO contributions should be dominant.  We find the results of our fits to be consistent with the expectations from chiral power-counting.  It would, nevertheless, be valuable to continue this study with the complete and correct NNLO formula including chiral logarithms.

The error in Table~\ref{tbl:error} labeled ``chiral-continuum extrapolation'' is estimated by performing a number of different ``reasonable" fits and taking the spread between them.  For example, we add additional higher-order analytic terms (giving fits with acceptable confidence levels) and compare the results with those obtained from the preferred fit.  We also vary the coefficient of the log terms between $f_0$ and $f_\pi$ as part of our estimate of the $f_\pi$ error, and between $f_\pi$ and $f_K$ as part of our estimate of the error in $f_K$.  Although we include terms proportional to $a^2$ in the preferred fit, there is some ambiguity (with only two lattice spacings) in the dominant source of discretization errors, which may be purely $a^2$ corrections, taste-breaking terms proportional to $\alpha_s^2 a^2$, or chiral symmetry breaking terms proportional to $m_{res} a^2$.  It is also possible that the different sources lead to discretization effects of the same size.  We therefore vary the coefficient of the NLO $a^2$ analytic terms and include the resulting spread in the decay constants as part of the systematic error.  We also include the parametric uncertainty coming from the uncertainty in the bare quark masses in the chiral extrapolation error.  



\begin{table}[h]
  \centering
  \caption{Preliminary error budget.  Uncertainties are shown as percentages.} \label{tbl:error}
  \begin{tabular}{ll@{\quad}c@{\quad}c@{\quad}c} \hline\hline
    source                     &$f_K$  &$f_\pi$  &$f_K/f_\pi$ \\ 
    \hline
    statistics                          & 1.1       & 1.5          & 1.3            \\  
    input $r_1$                      & 1.6       & 2.0          & 0.3            \\  
    chiral-continuum extrapolation          & 2.3       & 2.2          & 1.0         \\  
    finite volume                       & 0.3       & 0.9          & 0.9           \\  
    \hline
    total error                   & 3.0       & 3.4          & 1.9           \\  
    \hline\hline
  \end{tabular}

   \vspace{-7mm}
\end{table}

\section{Results and Conclusions}

Our preliminary results for the light pseudoscalar meson decay constants and their ratio are
\bea  f_\pi = 129.1(19)(40) \  \textrm{MeV}, \ \ \
        f_K = 153.9(17)(44) \  \textrm{MeV},  \ \ \
        f_K/f_\pi = 1.191(16)(17), \nonumber
\eea
where the first error is statistical and the second is the sum of systematic errors added in quadrature.  A breakdown of the total error for each quantity is given in Table~\ref{tbl:error}.  Our errors are somewhat reduced as compared to the results presented in the talk, due to the completion of an additional fine ensemble with a light strange sea quark.  Given the experimental values for the kaon and pion leptonic branching fractions and the electroweak corrections \cite{Amsler:2008zz}, we obtain $|V_{us}|/|V_{ud}|=0.2315(45)(7)$, where the first error is the lattice error (with statistical and systematic errors added in quadrature), and the second is the combined error from experiment and electroweak radiative corrections.  Taking the PDG value of $|V_{ud}|=0.97418(27)$ from superallowed $\beta$-decay \cite{Amsler:2008zz}, we obtain $|V_{us}|=0.2255(44)(7)$.  This is consistent with the value of $|V_{us}|=0.2255(19)$ coming from semileptonic kaon decays and non-lattice theory \cite{Amsler:2008zz}.  Given the above value of $|V_{ud}|$, our value of $|V_{us}|$ is also consistent with unitarity.

\end{document}